\documentclass[prl,preprint,floatfix,showpacs,preprintnumbers,superscriptaddress,amsmath,amssymb]{revtex4}
\usepackage[dvips]{graphics}
\usepackage{psfrag}
\usepackage{graphicx}
\usepackage{ucs}
\usepackage[utf8x]{inputenc}

\begin{document}

\title{Cross-correlations  in  transport  through  parallel  quantum  dots}

\author {Sebastian Haupt}
\affiliation{Forschungszentrum Karlsruhe, Institut f\"ur Nanotechnologie,
76021 Karlsruhe, Germany}
\affiliation{Institut f\"ur Theoretische Festk\"orperphysik, and DFG-Center for Functional Nanostructures (CFN),
Universit\"at Karlsruhe, 76128 Karlsruhe, Germany}

\author {Jasmin Aghassi}
\affiliation{Forschungszentrum Karlsruhe, Institut f\"ur Nanotechnologie,
76021 Karlsruhe, Germany}
\affiliation{Institut f\"ur Theoretische Festk\"orperphysik, and DFG-Center for Functional Nanostructures (CFN),
Universit\"at Karlsruhe, 76128 Karlsruhe, Germany}

\author {Matthias H. Hettler}
\affiliation{Forschungszentrum Karlsruhe, Institut f\"ur Nanotechnologie,
76021 Karlsruhe, Germany}

\author {Gerd Sch\"on}
\affiliation{Forschungszentrum Karlsruhe, Institut f\"ur Nanotechnologie,
76021 Karlsruhe, Germany}
\affiliation{Institut f\"ur Theoretische Festk\"orperphysik, and DFG-Center for Functional Nanostructures (CFN),
Universit\"at Karlsruhe, 76128 Karlsruhe, Germany}


\begin{abstract}
We investigate cross-correlations in the tunneling currents through two parallel quantum dots coupled to independent electrodes and gates and interacting via an inter-dot Coulomb interaction. The correlations reveal additional information, beyond what can be learned from the current or conductance, about the dynamics of transport processes of the system. We find qualitatively different scenarios for the dependence of the cross-correlations on the two gate voltages. Reducing the temperature below the inter-dot Coulomb interaction, regions of a given sign change from spherical shapes to angular L-shapes or stripes. Similar shapes have been observed in recent experiments. 
\end{abstract}

\maketitle
{\bf Introduction} -- 
The analysis of shot noise and current correlations in quantum dots reveals important information about the dynamics of the transport processes in the system, more than what can be obtained from the current or 
conductance.~\cite{blanter}
Due to the presence of extrinsic noise sources (e.g.\ $1/f$-noise) 
experiments are difficult to perform, but very recently two groups succeeded in measuring 
the so-called cross-correlations between two currents flowing through a quantum dot system.~\cite{DiCarlo,zhang,ensslin}
Cross-correlations are also important when 
a quantum dot setup is used as a beam-splitter.~\cite{Cottet_prl,Cottet_prb}

In this letter we study the cross-correlations for a model of the double dot system studied
by McClure {\it et al.} in Ref. ~\onlinecite{DiCarlo}. Two parallel quantum dots (top and bottom), interacting via an inter-dot Coulomb interaction, are coupled to four independent source or drain electrodes and two independent gates.
The cross-correlations between the currents through the two dots, plotted as a function of two gate voltages at fixed transport bias, form regions of positive or negative sign. Depending on the strength of the inter-dot Coulomb interaction, the temperature, and the transport bias we observe various shapes of these regions, namely spherical shapes, L-shapes, or stripes. In particular,  L-shapes and stripes develop as the temperature is lowered, confirming what may have been observed in experiments.~\cite{DiCarlo_private}
We find that the observed structure of the cross-correlations is due to the combination of various transport processes between the charge states that can be reached at the given transport bias.

\begin{figure}[t]
\centerline{\includegraphics[width=8.5cm]{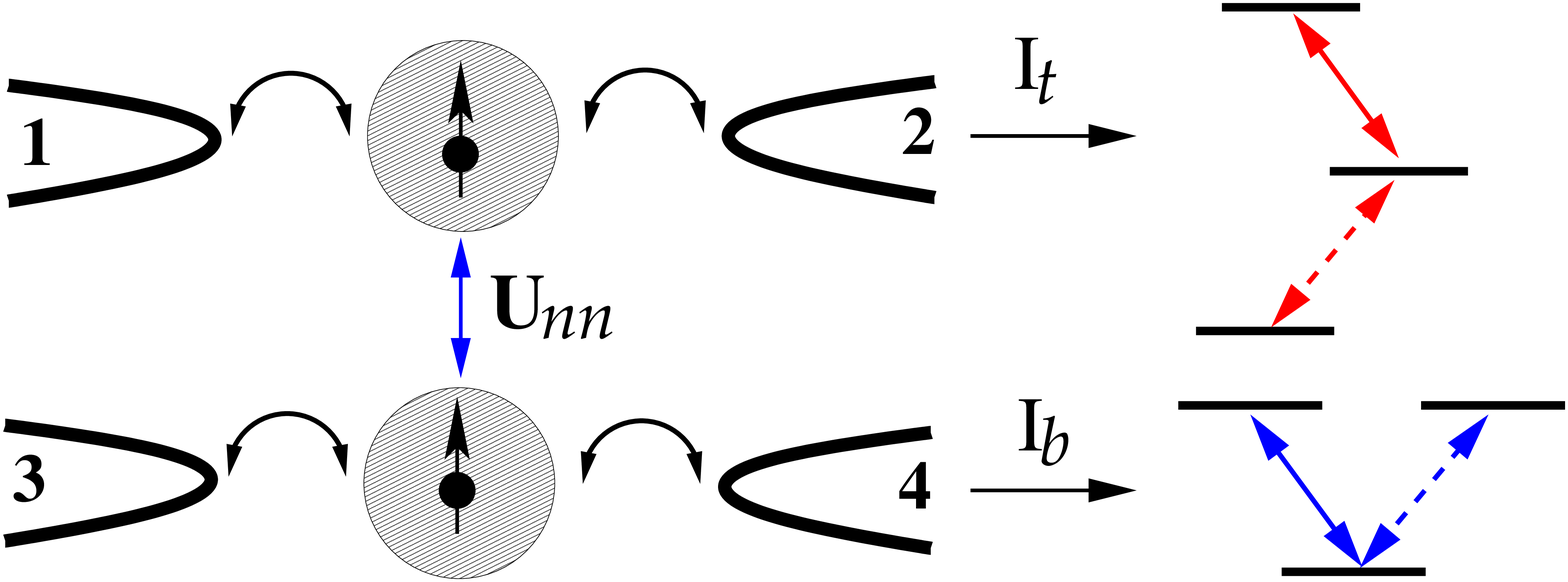}}
\caption{{\it Left panel: }sketch of the system. Two parallel quantum dots are coupled to four electrodes, interacting via the inter-dot Coulomb repulsion $U_{nn}$.
(Color online) {\it Right panel}: sketch of two competing/supporting processes which lead to
a different sign of the  cross-correlations. The dashed (solid) arrow indicates transport over the bottom (top) dot.} 
\label{fig:sketch}
\end{figure}

{\bf Model} -- 
We consider two parallel single-level quantum dots (top  and bottom ($i={\rm t,b}$)) with level energies  $\epsilon_{i}$, onsite Coulomb repulsion $U$ and inter-dot Coulomb interaction $U_{nn}$.
The double dot Hamiltonian reads     
\begin{equation}
\hat H_{\rm D} = \sum_{i \sigma} \epsilon_{i} n_{i \sigma} +  U\sum_i n_{i\uparrow}n_{i\downarrow} +  U_{nn}\sum_{<ij> \sigma \sigma'} n_{i\sigma}n_{j\sigma'}\nonumber \,.  
\label{hamilton} 
\end{equation}
Since tunneling between the dots is suppressed the energies of the double dot eigenstates are fully determined by the dot occupation 
number $n_{i\sigma}= c_{i\sigma}^\dagger c_{i\sigma}$.
The dots are coupled via the usual terms 
$H_{{\rm T},r}= \sum_{i k \sigma} \left( t^r_i a_{k \sigma r}^{\dag} c_{i\sigma} + h.c.\right)$ to four independent source and drain electrodes 
($r=1..,4$, see left panel of Fig.~\ref{fig:sketch}). The electrodes
(with operators $a_{k \sigma r}$)  are described by non-interacting electrons with density of states $\rho_r$ and chemical potential $\mu_r$. 
The tunneling amplitudes $t^r_i$ are chosen 
independent of electron spin $\sigma$ and lead momentum $k$.

The gate voltages $V_{\rm G,t}$ and $V_{\rm G,b}$ shift the energies of the double dot eigenstates by
$e V_{\rm G,t} \cdot (n_{\rm t} + \alpha\, n_{\rm b}) 
+ e V_{\rm G,b} \cdot (n_{\rm b} + \alpha \, n_{\rm t} )$
where $n_{\rm i}=\sum_\sigma n_{\rm i\sigma}$ are the occupation numbers of the dots $i={\rm t,b}$.  
The factor $\alpha$ is a measure for the unavoidable (in experiment) cross talk between gate voltages, 
which is of the order $\alpha=0.1$.~\cite{DiCarlo}
To compare to these experiments we choose the two left electrodes (and similar the two right ones) to be on the same potential $\mu_L, (\mu_R)$ and apply the transport bias $V_{\rm trans}=\mu_L -\mu_R$ symmetrically.

The current operators are defined by
$\hat{I}_r = -i(e/\hbar) \sum_{i k \sigma} \left(
 t^r_i a_{k \sigma r}^{\dag} c_{i\sigma} - h.c.\right)$.
Since there is no tunneling between the dots the currents over the top (bottom) channel are conserved. Therefore, we consider symmetric combinations over the corresponding source and drain electrodes, e.g.\ for current through the top dot  $\hat I_{t}=(\hat I_{1}-\hat I_{2})/2$ and $I_t = \langle \hat{I}_t \rangle$.
For each current the (zero-frequency) shot noise power spectrums is given by 
$S_{i} = \int_{-\infty}^{\infty} d\tau \langle \delta \hat{I}_i(\tau) \delta \hat I_i(0)
+ \delta\hat{I}_i(0)\delta \hat{I}_i(\tau)\rangle $
with $\delta\hat I_i(\tau)=\hat I_i(\tau)-\langle\hat I_i\rangle$. 
The cross-correlations $S_{tb}$ between top and bottom currents are defined as 
\begin{equation}
 S_{\rm tb} = \int_{-\infty}^{\infty} d\tau
\langle\delta \hat{I}_{\rm t}(\tau)\delta\hat I_{\rm b} (0)
+ \delta\hat{I}_{\rm t}(0)\delta\hat I_{\rm b}(\tau)\rangle\,  . 
\label{eq:crossdef}
\end{equation}

Expressions for the shot noise and and cross correlations in the sequential tunneling limit were first derived by Korotkov (Ref.~\onlinecite{korotkov} and references therein) within a master rate equation approach for single electron transistors, and by Cottet et al.~\cite{Cottet_prl,Cottet_prb} for several quantum dot models.
We can analyze the noise and cross-correlations in the frame of the real-time diagrammatic expansion~\cite{diagrams} in the coupling constants $\Gamma_r = 2\pi | t^r|^2 \rho_r$ to the leads.~\cite{thielmann_prb,thielmann_prl}
This is useful if higher order terms such as co-tunneling are investigated. In the present problem the main features are found in lowest order, i.e., for sequential tunneling already.
To describe this situation we can use Fermi's golden rule to obtain the transition rates ${W}_{\chi,\chi'}$ 
between eigenstates $|\chi\rangle$ and $|\chi'\rangle$ of the quantum dot system. Considering the processes involving the electrode $r$ we have 
\begin{eqnarray}
& W_{\chi',\chi}^{r}=  2\pi \rho_r & \sum_{\sigma} 
 f_{r}(E_{\chi',\chi})
 \left| \sum_i t^r_i \langle \chi'| c_{i\sigma}^{\dag} | \chi \rangle \right|^2 +
 \nonumber \\
& \,  &[1 - f_{r}(-E_{\chi',\chi})]
   \left| \sum_i t^r_i \langle \chi' | c_{i\sigma} | \chi \rangle \right|^2 
\end{eqnarray}
for $\chi' \neq \chi$, together with $W_{\chi,\chi}^{r} = 
- \sum_{\chi' \neq \chi} W_{\chi',\chi}^{r}$.
Here $E_{\chi',\chi}= E_{\chi'}-E_{\chi}$ is the energy difference
between the states involved,  $f(x)= 1/(\exp{(x/T)} +1)$ is the
Fermi function, 
and $f_{r}(x)=f(x - \mu_r)$. 
The total rate is the sum over all electrodes, ${\bf W} = \sum_r {\bf W}^{r}$
(bold face symbols indicate matrix/vector notation).
From the transition rates we obtain the
stationary occupation probabilities ${\bf p}^{st}$ from the 
stationary master equation, ${\bf W}{\bf p}^{st}=0$. 
To compute the current we also need to account for the
direction of tunneling process to/from the quantum dots and the considered lead.
The corresponding current rate, e.g., for the top current is given by
$W_{\chi,\chi'}^{I_{\rm t}}=(W_{\chi,\chi'}^{1}-W_{\chi,\chi'}^{2})
(\Theta(N_{\chi}-N_{\chi'})-\Theta(N_{\chi'}-N_{\chi}))$, 
where $N_{\chi}$ is the total number of electrons in the state $|\chi\rangle$.
The currents can then be expressed as 
$I_{{\rm t,b}} = {e\over 2\hbar} {\bf e}^T {\bf W}^{I_{{\rm t,b}}} {\bf p}^{{\rm st}}$, 
where for the efficiency of the notation we introduced 
${\bf e}^T =(1,\ldots ,1)$.~\cite{thielmann_prb}

In sequential tunneling, the expression for the shot noise (auto-correlation of a given current)
consists of two terms. One comes from the ``self-correlation'' of the two current operators, the other 
is due to the time evolution of the quantum dot system (in the presence of the electrodes) 
in the time between the action of current operators. For the cross-correlations, the first term vanishes in 
sequential tunneling.~\cite{korotkov,Cottet_prl,Cottet_prb} Therefore, the cross-correlations are concisely written as 
\begin{equation}
  S_{\rm tb}= \frac{e^2}{\hbar} {\bf e}^T {\bf W}^{I_{\rm t}}
    {\bf P} {\bf W}^{I_{\rm b}} {\bf p}^{{\rm st}} .
\label{eq:cross_seq}
\end{equation}
Here, the conditional probability ${\bf P}$ accounts for the time evolution mentioned above.~\cite{korotkov} It is determined via
${\bf W P} = {\bf p}^{\rm st} \otimes {\bf e}^T - {\bf 1}$, which can be derived from
the Dyson equation for the full time propagator of the quantum dot system.~\cite{thielmann_prb}
The more complicated expressions for the cross-correlations in higher order perturbation theory are a straightforward generalization~\cite{haupt-diplom} of the shot noise expressions, though care is necessary to correctly account for the rates due to the various electrodes. 
\begin{figure}[t]
\centerline{\hspace*{-0.5cm}\includegraphics[width=8.5cm]{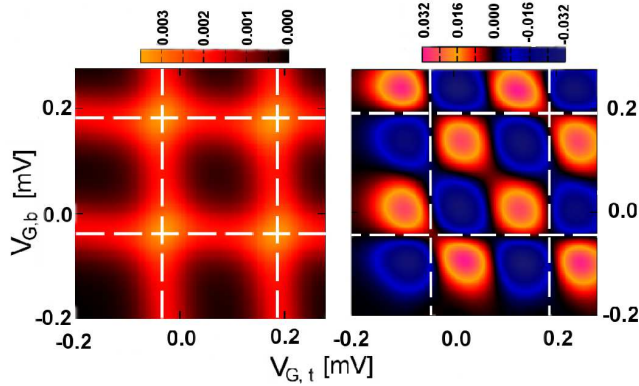}}
\caption{(Color online) {\it Left panel}: 
Stability diagram. Zero bias conductance vs. both gate voltages $V_{\rm G,t}$ and $V_{\rm G,b}$. The dashed lines separate the ground states
with different occupation number pairs. Between the four degeneracy vertices  
the ground state occupation is $(1,1)$. 
   {\it Right panel}: Cross-correlations vs. both gate voltages at a transport
voltage $V_{\rm trans}=0.1 {\rm mV}$. Red (bright) regions indicate positive and blue (dark) regions negative cross-correlations. 
 } 
\label{fig:stability}
\end{figure}

{\bf Results} --
Transport in quantum dot systems with a large charging energy $U$ and weak coupling $\Gamma_r$ is dominated by sequential processes, in which a quantum dot is charged by a single excess 
electron at a time. The transport voltage needs to overcome the sequential threshold  
for the single-electron charging, otherwise the transport is suppressed (Coulomb blockade). 
For the coupled dot system considered the sequential threshold 
of each dot depends on the gate voltages, and because of the inter-dot Coulomb repulsion $U_{nn}$ on the occupation of the other dot. 
The interactions between the dots may result in currents which mutually support or suppress  each other, leading to positive or negative cross-correlations $S_{\rm tb}$. 

For zero gate voltages, we consider dots with identical level energies 
$\epsilon_{\rm t}=\epsilon_{\rm b}=-0.2$ 
and intra-dot Coulomb repulsion $U = 0.2$ (all energies in meV). The inter-dot repulsion $U_{nn}=0.02$ is on the order of the temperature $k_{\rm B} T = 0.024$ for Fig.~\ref{fig:stability}. 
The dot-electrode couplings are chosen equal, $\Gamma_r=0.003 = k_{\rm B}T/8$.
This defines the line width of a dot level as $\Gamma = 2 \Gamma_r= 0.006$.


The stability diagram of the system, the differential conductance $dI/dV$ (in units of $\Gamma e^2/\hbar$)
of the total current $I=I_{\rm t}+I_{\rm b}$ at zero transport voltage  vs.\ both gate voltages, is shown in the left panel of Fig.~\ref{fig:stability}. 
The lines of non-zero conductance separate the regions with ground states corresponding to pairs of occupation numbers $(\langle n_{\rm t}\rangle,\langle n_{\rm b}\rangle)$, and form a (slightly) tilted square lattice  (honeycomb lattice, if inter-dot tunneling would be included). 
The intersection points (vertices) of the lines correspond the situation when 
ground states of four occupation number pairs are degenerate. 
As for our single-level model the occupation is limited to at most two electrons per dot, 
we have nine possible occupation number pairs (with corresponding spin degeneracy) 
Between the four degeneracy vertices, the ground state has occupation $(\langle n_{\rm t}\rangle,\langle n_{\rm b}\rangle)$=$(1,1)$,  
i.e. each dot is occupied by one electron at zero bias.

We now apply a fixed transport voltage $V_{\rm trans}=0.1 {\rm mV}$ over each dot and plot the
cross-correlations (also in units of $\Gamma e^2/\hbar$) of the system vs.\ both gate voltages, see right panel of Fig.~\ref{fig:stability}. We find sixteen sectors with positive (red) and negative correlations (blue). The cross-correlations form a checkerboard pattern,  where four sectors form a four-leaved clover shape around a degeneracy vertex. Note that the sign of the cross-correlations is not simply a ground state property, as
there are e.g. four spherical regions of alternating sign in the area with ground state $(1,1)$. Such a clover structure around degeneracy vertices was observed in experiments~\cite{DiCarlo} with two parallel multi-level quantum dots. The different signs were explained within a minimal model of two competing/supporting processes.

Let us consider the clover structure at the bottom left degeneracy vertex of Fig.~\ref{fig:stability} (choosing another vertex would simply need a change of occupation numbers). In the area around zero gate voltages, $V_{\rm G,t}\approx V_{\rm G,b}\approx 0$, the ground state is $(\langle n_{\rm t}\rangle,\langle n_{\rm b}\rangle)$ =$(1,1)$ and there are two nearly degenerate excited states $(2,1)$ and $(1,2)$ that can be reached at the given transport bias. 
The two transport processes $(1,1)\longrightarrow  (1,2)$ and $(1,1)\longrightarrow  (2,1)$
correspond to currents over different dots, see right panel of Fig.~\ref{fig:sketch}, lower part.  
Both processes have the same probability but only one is possible at a given time,
as the realization of one process preempts the other.
Therefore, the two currents are competing and the resulting cross-correlations are negative. 

We now decrease the bottom gate voltage so that we reach the region of positive cross-correlations.
For gate voltages, $V_{\rm G,t}\approx0.0 {\rm mV}, \,  V_{\rm G,b}\approx - 0.1 {\rm mV}$,
the ground state is now $(1,2)$, the closest excited state is $(1,1)$. Transport via 
the lower dot, $(1,2)\longrightarrow  (1,1)$, now allows transport over the upper dot, via $(1,1)\longrightarrow  (2,1)$, which is still within reach at the given transport bias, see right panel of Fig.~\ref{fig:sketch}, upper part. Therefore, current over the lower dot supports the transport over the upper dot, and the cross-correlations are positive.

We ignored so far that because of the relatively small inter-dot 
interaction  $U_{nn}$ the state $(2,2)$ can also be reached  from  $(1,2)$ (or $(2,1)$)
in nearly the same gate voltage region than the states $(1,2)$ ($(2,1)$) can be reached from $(1,1)$. Therefore, the processes involving the state $(2,2)$ are also involved. 
However, the above reasoning still holds 
if the temperature $T$ is larger than $U_{nn}$, which was the case so far.
\begin{figure}[t]
\centerline{\hspace*{-0cm}\includegraphics[width=8.5cm]{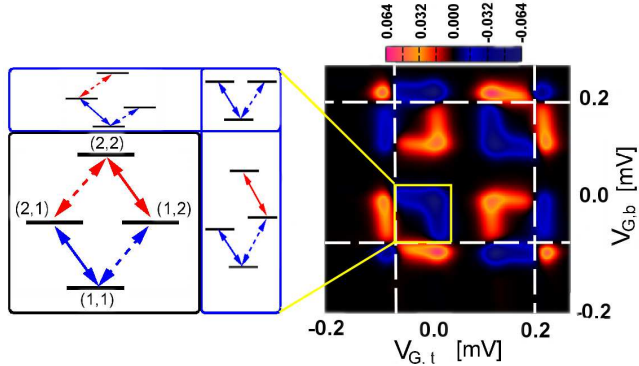}}
\caption{(Color online) {\it Left panel}: Sketch of the processes relevant for
the lower left negative $L-shape$. Different combinations of positive and negative processes result in the $L-shape$ of the cross-correlations. 
{\it Right panel}: 
Cross-correlations vs. both gate voltages $V_{\rm G,t}$ and $V_{\rm G,b}$ at a temperature $k_{\rm B}T=0.008$. The cross-correlations develop into
$L-shapes$ inside the central region with ground state $(1,1)$. 
The dashed lines separate the regions of different ground state occupation.}
\label{fig:L-shape}
\end{figure}

This changes dramatically when we decrease the  temperature, such that 
$U_{nn}=0.02 > k_{\rm B} T=0.008$ (we also lower the lead couplings to $\Gamma_r=0.001$). Now we observe a very different structure of the cross-correlations, as seen in the right panel of Fig. \ref{fig:L-shape}.
The symmetric clover shape around a degeneracy vertex changes to {\it L-shapes} 
in the region with the ground state $(1,1)$, and stripes/bubbles
for the other regions. 

The L-shape around zero gate voltages (ground state $(1,1)$) 
is due to a break up of the previously spherical region with negative cross-correlations into four regions, which can be distinguished by the number
of ways (if any) the state $(2,2)$ can be reached at the given bias.
As discussed above, around zero gate voltages the system has processes $ (1,1) \longrightarrow (1,2)$ and 
$ (1,1)  \longrightarrow (2,1)$ available, 
which leads to  negative cross-correlations. On the other hand, if the 
system can also reach the state  $(2,2)$ from $(2,1)$ and $(1,2)$, this would bring about positive cross-correlations. The value of the cross-correlations is a now a question of how much the ``positive'' processes can compensate the negative cross-correlations of the processes related to the ground state $(1,1)$.

If both gate voltages are positive, the state $(2,2)$ can not be reached at all,
and we are back to the earlier explanation of (maximum) negative cross-correlations.
If we decrease one of the gate voltages (but keep the other fixed) at some point the transport voltage becomes sufficient so that the system can make one of the
processes $ (1,2) \longrightarrow (2,2)$ or  $ (2,1) \longrightarrow (2,2)$, 
depending on which gate voltage was lowered. This additional ``positive'' process 
leads to less negative cross-correlations in the ``legs'' of the L-shape. 

If we decrease both gate voltages in the same way from zero gate voltage, going along the diagonal in 
Fig.~\ref{fig:L-shape}, we reach an area where the cross-correlations are practically vanishing.
Here, the state $(2,2)$ can be reached from both states  $(1,2)$  and  $ (2,1)$
and, in combination, the positive and negative processes effectively cancel for the cross-correlations. Nevertheless, a finite current flows over both dots.

The L-shapes with positive cross-correlations are explained by a similar combinations of processes, though in reverse order, such that the initially positive cross-correlations are more and more compensated
by negative processes as the gate voltages are changed.
That we observe no L-shapes for the regions not associated with the $(1,1)$ ground state is partly due to the limitation of our model to single-level quantum dots. Multi-level dots with a 
finite level spacing could induce even more structure than presently observed. On the other hand, the
present shapes are clearly a consequence of the four charge states around the degeneracy vertices. 
This structure should be fairly robust given a sufficiently large charging energy $U$.
\begin{figure}[t]
\centerline{\hspace*{-0.5cm}\includegraphics[width=8.5cm]{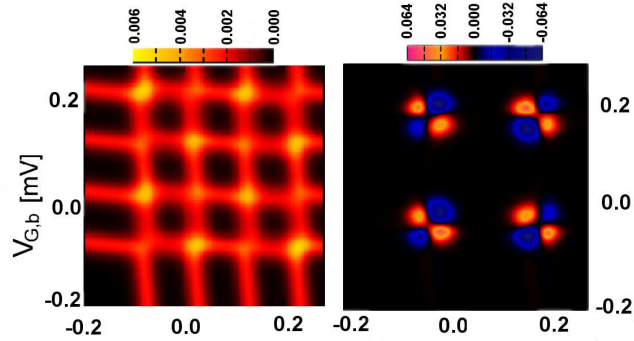}}
\caption{(Color online) {\it Left panel}: Differential conductance vs. both gate  voltages $V_{\rm G,t}$ and $V_{\rm G,b}$ at a fixed bias $V_{\rm trans}=0.1{\rm mV}$ and $k_{\rm B}T=0.008$. 
The conductance is enhanced in the areas with maximum positive cross-correlations. 
{\it Right panel}: Cross-correlations at a small bias $V_{\rm trans}=0.05{\rm mV}$ vs. both gate voltages. 
The  cross-correlations regions show clover structure and
are non-zero only near the degeneracy vertices.} 
 \label{fig:misc}
\end{figure}

As the transport bias dictates which states are available for sequential transport
its value will strongly influence the observed shape of the cross-correlations.
In the right panel of Fig.~\ref{fig:misc} we display the cross-correlations vs.\ both gate voltages
for a reduced bias voltage of  $V_{\rm trans}=0.05 {\rm mV}$. 
In this plot the L-shapes vanish and only clearly 
separated clover structures at the  degeneracy points are visible, similar to the cross-correlations for higher temperature. At larger transport bias (not shown) the regions with finite cross-correlations grow until they start to ``overlap''.

The left panel of Fig.~\ref{fig:misc}  depicts the differential conductance $dI/dV$ of the total current 
vs.\ both gate voltages for the fixed bias voltage $V_{\rm trans}=0.1 {\rm mV}$. The conductance has some similarity to the cross-correlations. Transport over each dot results in four vertical and horizontal lines of high conductance (orange-yellow), leading to sixteen intersection points. 
These intersection points are at same gate voltages than the points of maximum amplitude of the cross-correlations of Fig.~\ref{fig:L-shape}. In addition, intersection points with maximum positive cross-correlations correspond to maximum conductance, whereas points with negative cross-correlations show lower conductance. 
However, there are lines with high conductance where the cross-correlations essentially vanish. There, the transport is carried only over one dot, while the other dot is in the Coulomb blockade. 

{\bf Summary.} --
In summary, we found that the cross-correlations of currents through two parallel quantum dots are 
strongly influenced by the inter-dot Coulomb interactions, the temperature and the transport bias.
Processes favoring different signs of cross-correlations lead to various shapes,
similar to experimental findings.  
If the inter-dot Coulomb interaction is much larger than the temperature, the shape changes with increasing transport voltage from a four leaved clover around the charge degeneracy points, to L-shapes and stripes, before the regions start to overlap at large bias. 
In contrast to the stability diagram, 
the cross-correlations are not determined by the ground state alone, 
but depend sensitively on the excited states and the sequence of processes by which these states are 
reached.

{\em Acknowledgments.} We acknowledge stimulating discussions with Leonardo DiCarlo.

\end{document}